\documentclass[11pt]{article}
\usepackage{graphicx}
\usepackage[margin=1.25in]{geometry}
\usepackage[usenames,dvipsnames]{color}
\usepackage{url}
\usepackage[colorlinks = true,
            linkcolor = blue,
            urlcolor  = blue,
            citecolor = blue,
            anchorcolor = blue,
            hyperfootnotes=false]{hyperref}

\date{}

\def\Acknowledgements{\bigskip  \bigskip \begin{center} \begin{large}
             \bf ACKNOWLEDGEMENTS \end{large}\end{center}}

\def\Acknowledgements{\bigskip  \bigskip \begin{center} \begin{large}
             \bf ACKNOWLEDGEMENTS \end{large}\end{center}}

\def\beq{\begin{equation}}
\def\eeq#1{\label{#1}\end{equation}}
\def\eeqn{\end{equation}}

\newenvironment{Eqnarray}%
   {\arraycolsep 0.14em\begin{eqnarray}}{\end{eqnarray}}
\def\beqa{\begin{Eqnarray}}
\def\eeqa#1{\label{#1}\end{Eqnarray}}
\def\eeqan{\end{Eqnarray}}

\let\bar=\overbar

\def\lsim{\mathrel{\raise.3ex\hbox{$<$\kern-.75em\lower1ex\hbox{$\sim$}}}}
\def\gsim{\mathrel{\raise.3ex\hbox{$>$\kern-.75em\lower1ex\hbox{$\sim$}}}}

\def\del{\partial}
\def\Dslash{\not{\hbox{\kern-4pt $D$}}}
\def\dslash{\not{\hbox{\kern-2pt $\del$}}}
\def\pslash{\not{\hbox{\kern-2pt $p$}}}
\def\ETmiss{\not{\hbox{\kern-4pt $E$}}_T}

\def\Dlr{\mathrel{\raise1.5ex\hbox{$\leftrightarrow$\kern-1em\lower1.5ex\hbox{$D$}}}}

\def\mt{m_t}

\def\MSB{{\bar{M \kern -2pt S}}}
\def\msb{{\bar{\scriptsize M \kern -1pt S}}}

\def\drb{{\bar{\scriptsize D \kern -1pt R}}}

\def\GeV{{\rm GeV}}
\def\TeV{{\rm TeV}}

\newcommand\snowmass{\begin{center}\rule[-0.2in]{\hsize}{0.01in}\\\rule{\hsize}{0.01in}\\
\vskip 0.1in Submitted to the  Proceedings of the US Community Study\\ 
on the Future of Particle Physics (Snowmass 2021)\\ 
\rule{\hsize}{0.01in}\\\rule[+0.2in]{\hsize}{0.01in} \end{center}}

\usepackage{lineno}

\usepackage{ifthen}
\usepackage{fancyhdr}
\fancypagestyle{plain}{%
  \fancyhf{}%
  \fancyfoot[R]{\thepage}%
}

\usepackage{authblk}
\usepackage{epstopdf}
\usepackage{xcolor}
\usepackage[symbol]{footmisc}

\usepackage{xspace}
\newcommand\madgraph{\textsc{MadGraph5\_aMC@NLO}\xspace}
\newcommand\pythia{\textsc{Pythia8}\xspace}
\newcommand\delphes{\textsc{Delphes}\xspace}
\newcommand\rivet{\textsc{Rivet}\xspace}

\renewcommand\TeV{~\ensuremath{\textrm{TeV}}\xspace}
\renewcommand\GeV{~\ensuremath{\textrm{GeV}}\xspace}

\newcommand\subT{\ensuremath{_\mathrm{T}}}
\newcommand\pt{\ensuremath{p_\mathrm{T}}\xspace}
\renewcommand\mt{\ensuremath{m_\mathrm{T}}\xspace}

\newcommand{\vplusjets}{\ensuremath{V+\mathrm{jets}}\xspace}
\newcommand{\wplusjets}{\ensuremath{W+\mathrm{jets}}\xspace}
\newcommand{\zplusjets}{\ensuremath{Z+\mathrm{jets}}\xspace}

\newcommand{\ttbar}{\ensuremath{t\overline{t}}\xspace}

\begin{document}

\title{Background Monte Carlo Samples for a Future Hadron Collider}

\author[1]{Robert Gardner}
\author[2]{Simone Pagan Griso}
\author[3]{Stefan Hoeche}
\author[1]{Karol Krizka}
\author[4]{Fabio~Maltoni}
\author[5]{Andrew Melo}
\author[6]{Meenakshi Narain}
\author[7]{Isabel Ojalvo}
\author[1]{Pascal Paschos}
\author[8]{Laura Reina}
\author[9]{Michael Schmitt}
\author[10]{Horst Severini}
\author[11]{Giordon Stark}
\author[10]{John~Stupak~III\thanks{ john.stupak@ou.edu}}
\author[12]{Thiago Tomei}
\author[13]{Alessandro Tricoli}
\author[6]{David Yu}

\affil[1]{Enrico Fermi Institute, University of Chicago}
\affil[2]{Lawrence Berkeley National Laboratory}
\affil[3]{Fermi National Accelerator Laboratory}
\affil[4]{Centre for Cosmology, Particle Physics and Phenomenology, Université Catholique de Louvain}
\affil[5]{Department of Physics, Vanderbilt University}
\affil[6]{Department of Physics, Brown University}
\affil[7]{Department of Physics, Princeton University}
\affil[8]{Department of Physics, Florida State University}
\affil[9]{Department of Physics, Northwestern University}
\affil[10]{Homer L. Dodge Department of Physics \& Astronomy, University of Oklahoma}
\affil[11]{Santa Cruz Institute for Particle Physics, University of California, Santa Cruz}
\affil[12]{Department of Physics, Sao Paulo State University}
\affil[13]{Brookhaven National Laboratory}

\thispagestyle{fancy}
\fancyhf{}

\maketitle
\vspace{-0.75in}
\snowmass
\vspace{-0.25in}

\begin{abstract}
    A description of Standard Model background Monte Carlo samples produced for studies related to future hadron colliders.
\end{abstract}

\def\thefootnote{\fnsymbol{footnote}}

\setcounter{footnote}{0}


\section{Introduction}

The final missing piece of the Standard Model (SM), the Higgs boson, was discovered in 2012 by the ATLAS and CMS experiments at the Large Hadron Collider (LHC)~\cite{201230,20121}.  However, the SM leaves several important questions unanswered, whose explanation requires Beyond the SM (BSM) physics.  Despite the remarkable success of the LHC physics program, no statistically significant evidence for BSM physics has been observed.

The discovery of BSM physics may be just around the corner at the LHC.  Alternatively, the mass scale of BSM physics may be outside the reach of the LHC, requiring a future high-energy collider to uncover.  A  $\sqrt{s}=100$~TeV $pp$ collider has therefore been proposed to eventually supplant the LHC~\cite{FCC:2018vvp}.

The cross sections for SM background processes such as $V(W^{\pm}/Z)$+jets and $t\bar{t}$ are very large at $\sqrt{s}=100$~TeV.  Accurately modeling the tails of kinematic distributions for these processes requires production of sophisticated Monte Carlo (MC) samples with large statistics.  The production of such samples, following an approach similar to that adopted for Snowmass 2013~\cite{snow2013a,snow2013b,snow2013c}, is described here.

\section{Monte Carlo Simulation}

High-statistic \vplusjets and \ttbar background Monte Carlo samples were produced for $\sqrt{s}=13$ and 100 TeV $pp$ colliders, using \madgraph v3.3.1~\cite{Alwall_2014}, \pythia v8.306~\cite{pythia}, and \delphes v3.5.0~\cite{de_Favereau_2014}, with computing resources provided by the Open Science Grid~\cite{osg07}.  \madgraph was used to simulate $pp$ hard scatter matrix elements.  Parton shower and hadronization were performed with \pythia.  \delphes parameterized detector simulation was used to account for the detector response.

\subsection{Hard Scatter}

Hard scatter matrix elements were computed in the four-flavor scheme at leading-order (LO) accuracy in the strong coupling constant with \madgraph.  On-shell heavy resonances ($V, t, H$) were treated as stable particles.  Radiated partons are allowed (with transverse momentum $\pt\geq20\GeV$ and pseudorapidity $|\eta|<5$), up to a total of four final state particles.  The \texttt{NNPDF31\_nnlo\_as\_0118} parton distribution function was used~\cite{nnpdf}.

So-called ``gridpacks'' were produced for streamlined operation on the OSG computing grid using the \texttt{makeGridpacks.sh} script, contained within the \texttt{MCProd} package (git tag: \texttt{v1.2nobias}).\footnote{\href{https://github.com/Snowmass21-software/MCProd/releases/tag/v1.2nobias}{https://github.com/Snowmass21-software/MCProd/releases/tag/v1.2nobias}}  Configuration cards can be found in the \texttt{Cards} directory.


\subsection{Parton Shower}
On-shell heavy resonances were decayed using \pythia, neglecting spin correlation effects.  Parton shower and hadronization were also simulated with \pythia, using the CMS \texttt{CP5} tune~\cite{Khachatryan_2016}.  $k\subT$-MLM matching and merging of radiation from the matrix-element and parton shower was performed using the \textsc{MG5aMC\_PY8\_interface}~\cite{Alwall:2007fs}.  For the \vplusjets (\ttbar) samples, a matching scale $\texttt{XQCUT}=40$ (80)\GeV was used.  In both cases, \texttt{QCUT} is taken to be $1.5\times\texttt{XQCUT}$.

\subsection{Detector Response}

The detector response was simulated using \delphes.  The parameterized detector performance is based on test beam data and full-simulation of a future FCC-hh detector~\cite{Selvaggi:2717698}.  The effect of pileup interactions is not simulated explicitly; instead, the parameterized efficiency and resolution for various physics objects include the expected degradation due to pileup interactions.

\subsection{Workflow}

\begin{figure}[h]
\centering
\includegraphics[width=8cm]{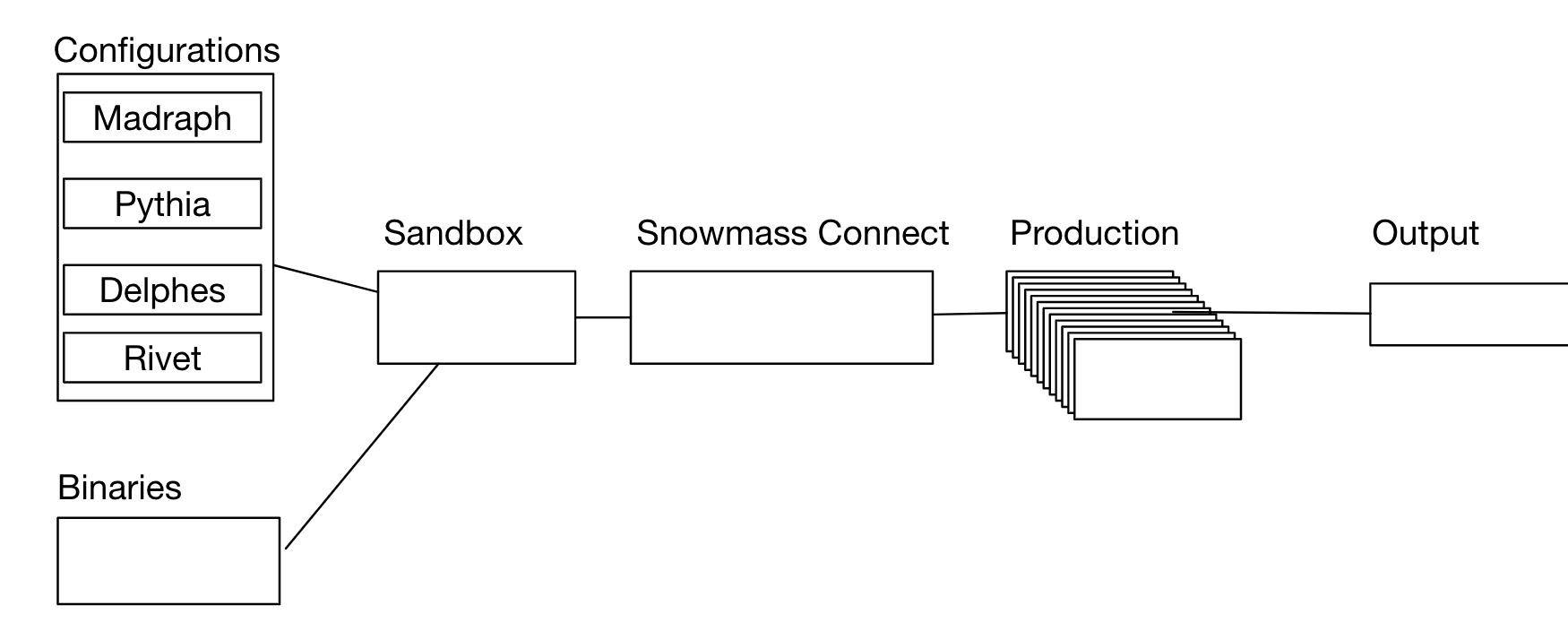}
\includegraphics[width=8cm]{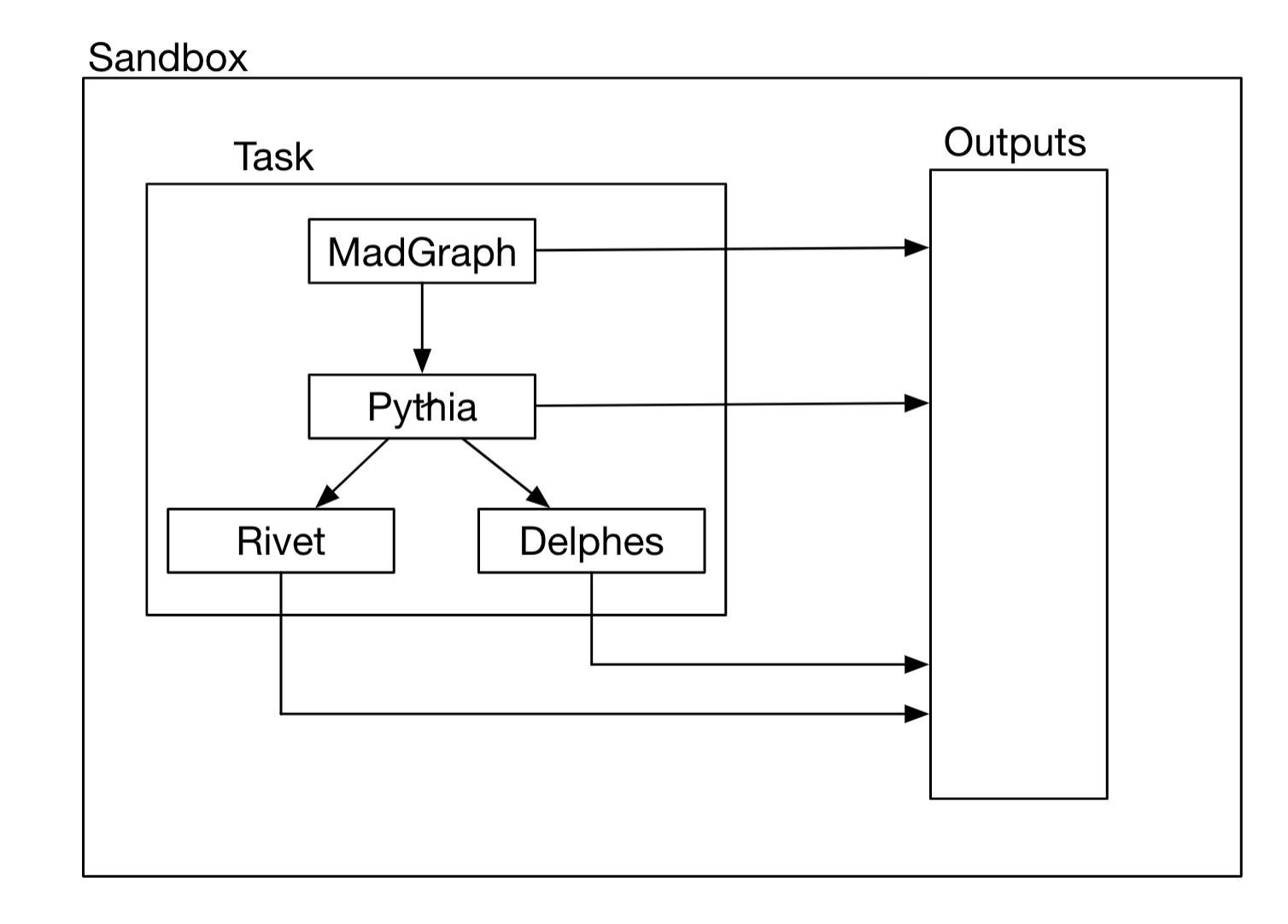}
\caption{A summary of the production workflow.}
\label{fig:pipe}       
\end{figure}

A summary of the production workflow is shown in Fig.~\ref{fig:pipe}.  The \madgraph output (a standard \texttt{LHE} file~\cite{Alwall:2006yp}) serves as input to \pythia.  The \pythia output (in \texttt{HepMC} format~\cite{Buckley:2019xhk}) serves as input to both \rivet and \delphes.  \delphes produces a standard \texttt{ROOT} output~\cite{rene_brun_2019_3895860}, while \rivet produces output in the \texttt{YODA} format~\cite{rivet}.  In addition to storing the final outputs from this workflow, intermediate outputs are retained as well.

\section{Validation}

Validation of the matching/merging of phase space for parton emission covered by the matrix element and parton shower is performed by studying differential jet rates.  Comparison of the $\sqrt{s}=13$\TeV samples to unfolded LHC data was performed using \rivet~\cite{rivet}.

\subsection{Matching/Merging}

Suitability of the matching/merging settings is confirmed based on the differential jet rates, shown for $\sqrt{s}=100$\TeV \wplusjets events in Figure~\ref{fig:djrs}.  Similar plots for \zplusjets and \ttbar samples are shown in the appendix.  These observables represent the scale at which an $N$-jet event transitions to an $N+1$-jet event, and exhibit a smooth transition. 

\begin{figure}
    \begin{center}
        \includegraphics[width=0.48\textwidth,page=1,trim=0.9in 4in 0.6in 1.25in,clip]{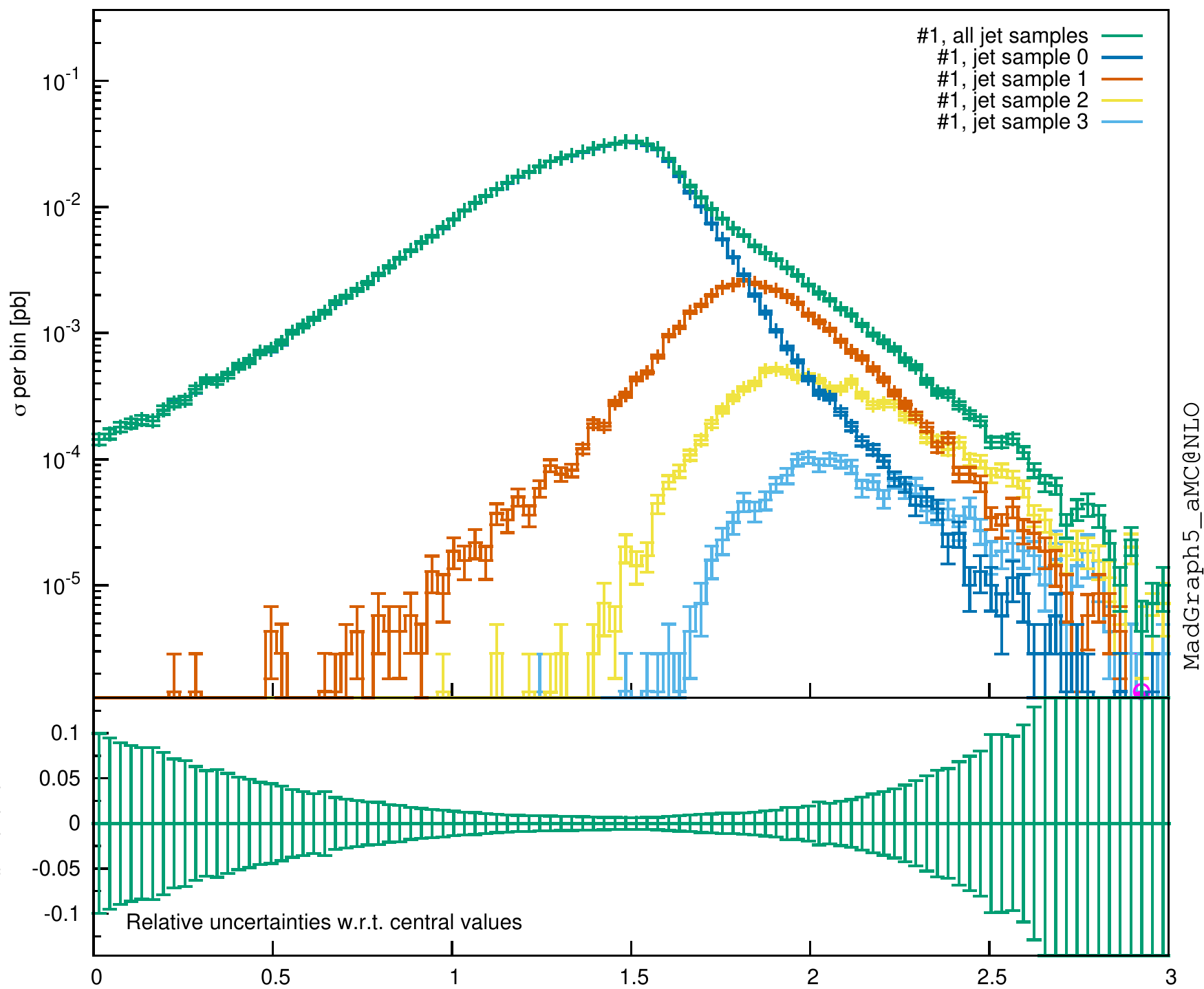}
        \includegraphics[width=0.48\textwidth,page=2,trim=0.9in 4in 0.6in 1.25in,clip]{img/100TeV_W.tar.gz.pdf}
        \includegraphics[width=0.48\textwidth,page=3,trim=0.9in 4in 0.6in 1.25in,clip]{img/100TeV_W.tar.gz.pdf}
        \includegraphics[width=0.48\textwidth,page=4,trim=0.9in 4in 0.6in 1.25in,clip]{img/100TeV_W.tar.gz.pdf}
        \caption{Differential jet rate distributions for $\sqrt{s}=100$\TeV \wplusjets events, showing log$_{10}$ of the merging scale.  The upper left  (right) plot represents a transition from a 0-jet event to a 1-jet (1-jet event to a 2-jet) event.  The lower left (right) plot represents a transition from a 2-jet event to a 3-jet  (3-jet event to a 4-jet) event.}
        \label{fig:djrs}
	\end{center}
\end{figure}	

\subsection{Comparison to Data}

The \rivet toolkit provides a convenient mechanism to compare simulated MC to unfolded collider data, using a variety of predefined analysis routines.  

The \wplusjets sample was validated using the \texttt{CMS\_2017\_I1610623} routine, which targets $W(\rightarrow\mu\nu)$+jets production.  This routine selects events with a muon and transverse mass $\mt\geq50\GeV$.  Muons are required to have $\pt\geq25\GeV$ and $|\eta|<2.4$.

The \zplusjets sample was validated using the \texttt{CMS\_2019\_I1753680} routine, which targets $Z(\rightarrow\ell\ell)$+jets production, where $\ell=e/\mu$.  This routine selects events with a pair of opposite-sign electrons or muons.  In both cases, electrons and muons are required to have $\pt\geq25\GeV$ and $|\eta|<2.4$.

The \ttbar sample was validated using the \texttt{CMS\_2018\_I1663958} routine, which targets events with semi-leptonic decay of top pairs.  This routine selects events with an electron or muon (with $\pt\geq30\GeV$ and $|\eta|<2.4$).  $W$ and $t$ candidates are formed from the lepton, jets and missing transverse energy $\ETmiss$, by minimizing the difference between the true mass and the reconstructed mass for these candidates.

\section{Results}

Cross sections for $\sqrt{s} = 13$ and 100\TeV background processes are given in Table~\ref{tab:sigma}.

\begin{table}[h!]
    \centering
    \begin{tabular}{|c | c c c|} 
        \hline
        & \wplusjets & \zplusjets & \ttbar \\\hline
        $\sigma_{13\TeV}$ [nb] & 184.7 & 56.05 & 0.6137 \\
        $\sigma_{100\TeV}$ [nb] & 1,428 & 471.1& 36.47\\
        \hline
    \end{tabular}
    \caption{Calculated cross sections for SM background processes at $\sqrt{s}=13$ and 100\TeV.}
    \label{tab:sigma}
\end{table}

\begin{figure}
    \begin{center}
        \includegraphics[width=0.48\textwidth]{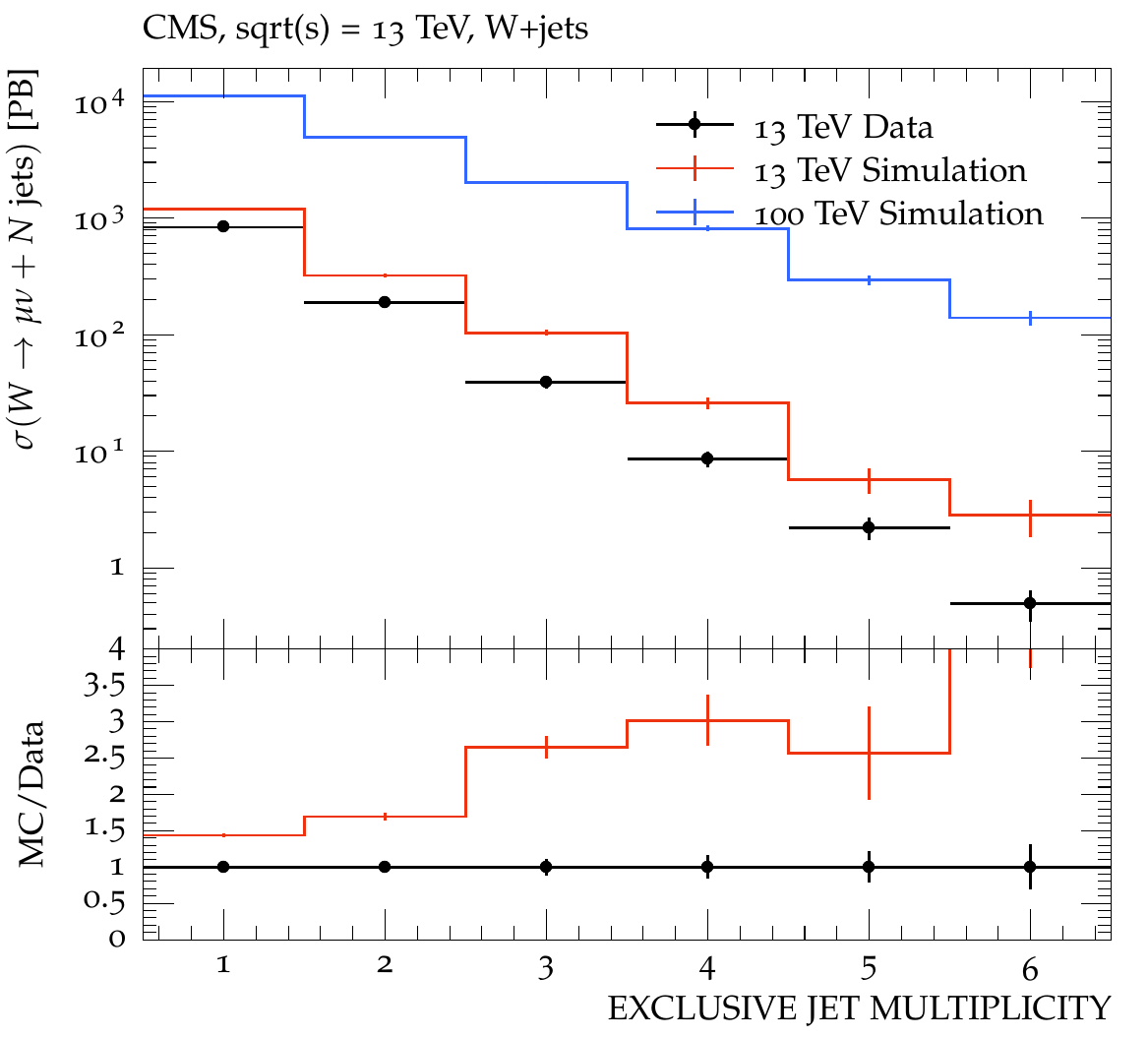}
        \includegraphics[width=0.48\textwidth]{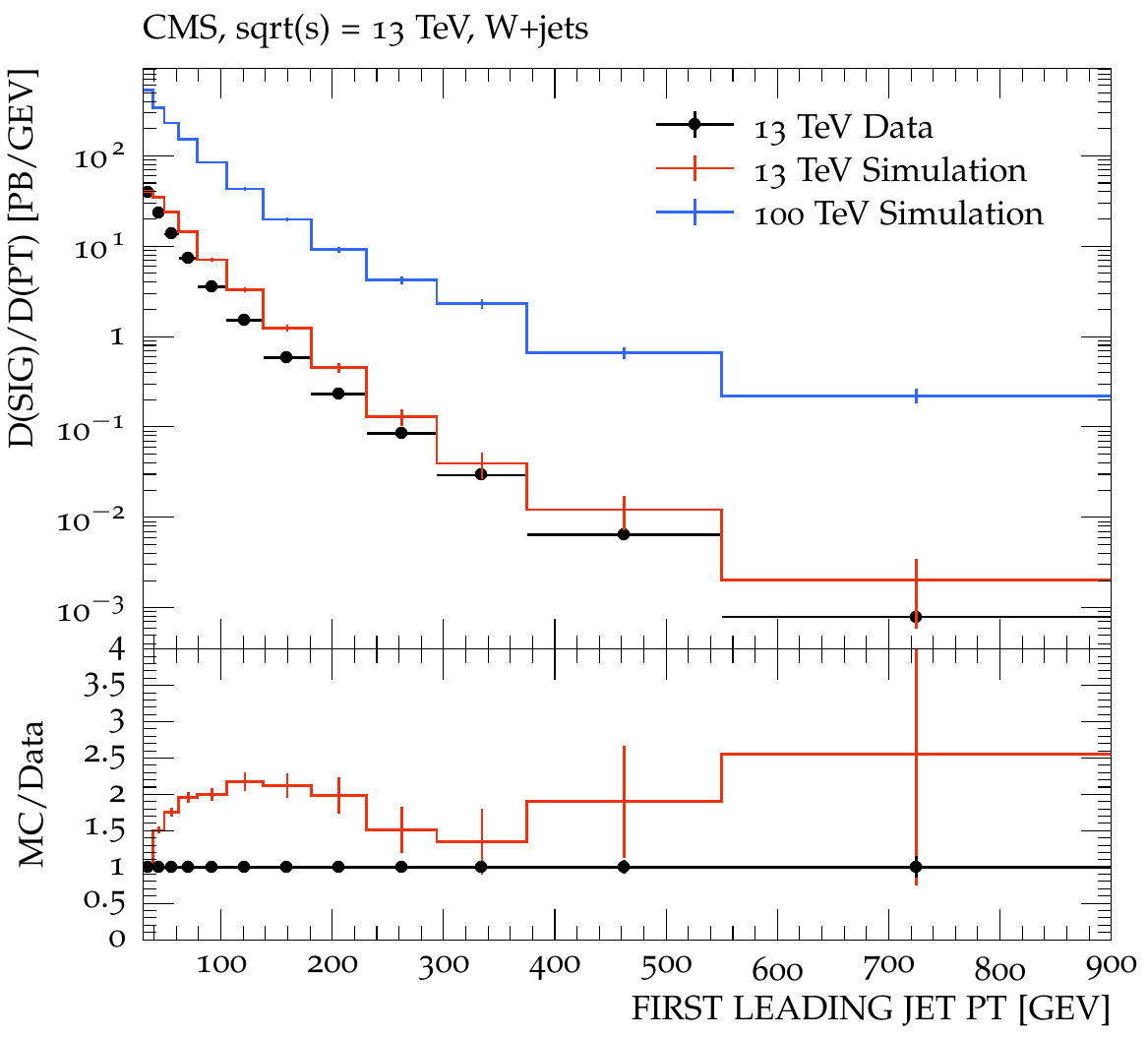}
        \includegraphics[width=0.48\textwidth]{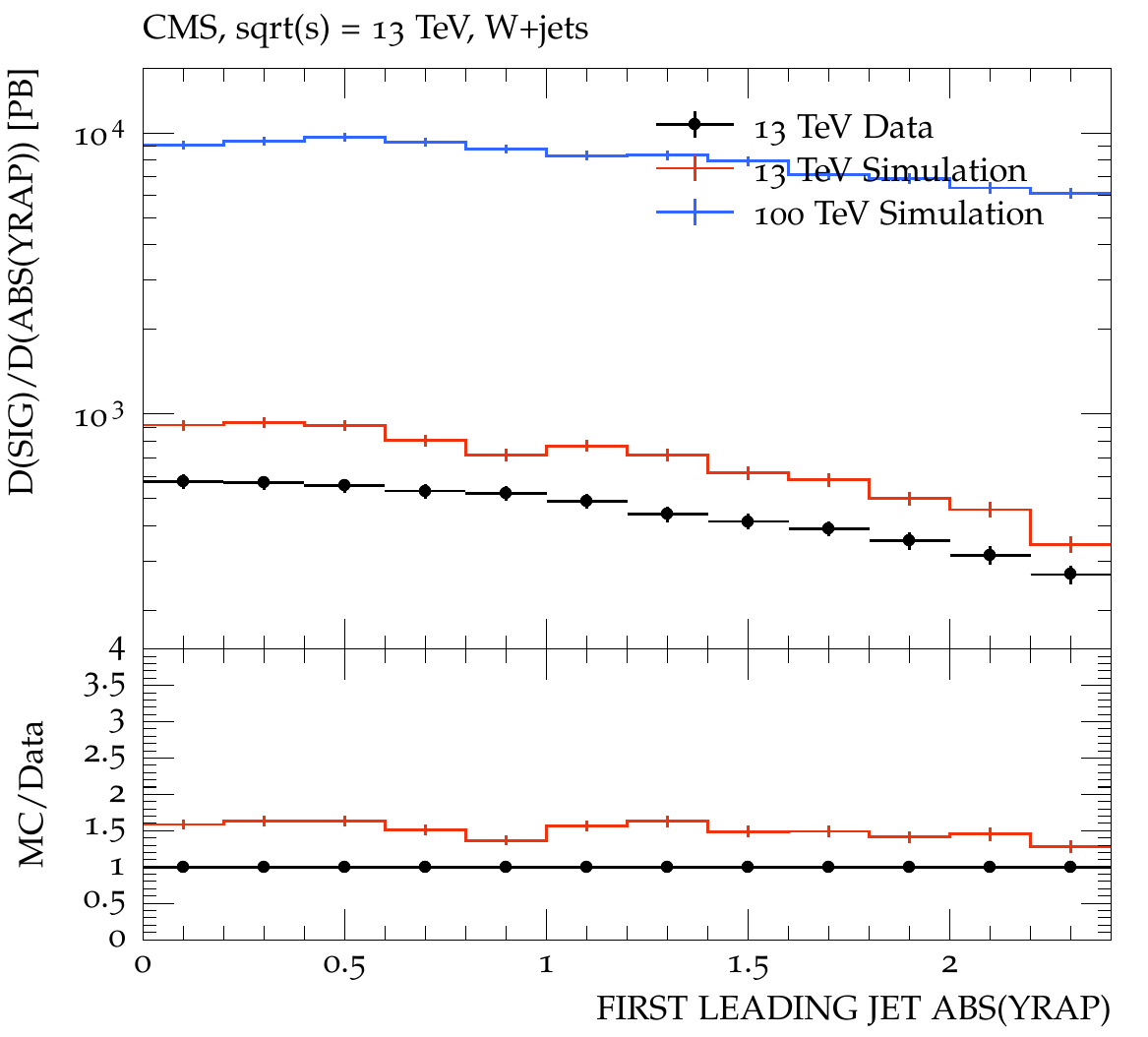}
        \includegraphics[width=0.48\textwidth]{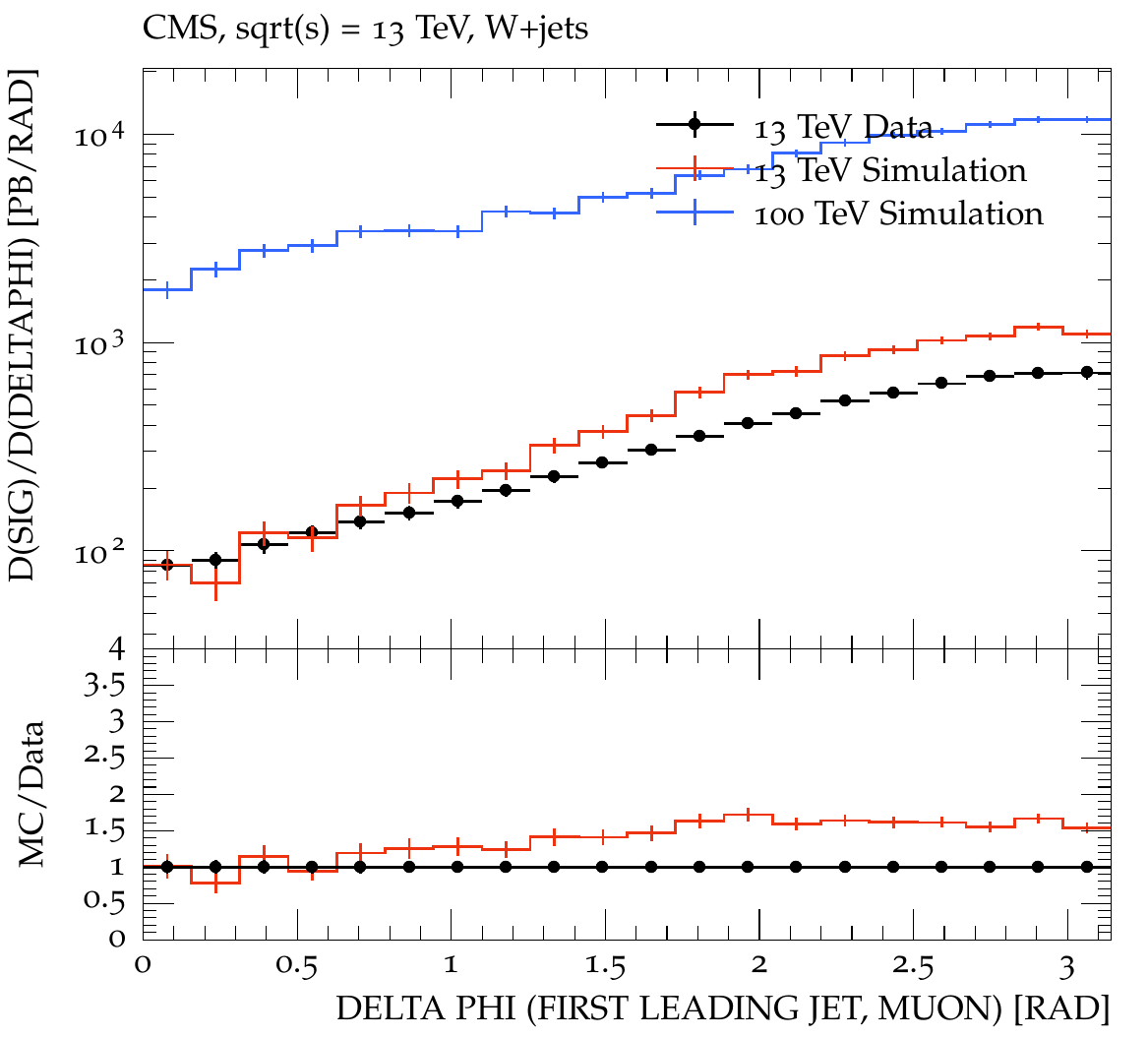}
        \caption{A comparison of $\sqrt{s}=13$\TeV \wplusjets data and simulation, as well as results for $\sqrt{s}=100$\TeV simulation.  The upper left (right) plot shows the exclusive jet multiplicity (leading jet $p_T$).  The lower left (right) plot shows the leading jet rapidity $y$ (difference in azimuthal angle between the selected muon and leading jet).}
        \label{fig:W}
    \end{center}
\end{figure}

\begin{figure}
    \begin{center}
        \includegraphics[width=0.48\textwidth]{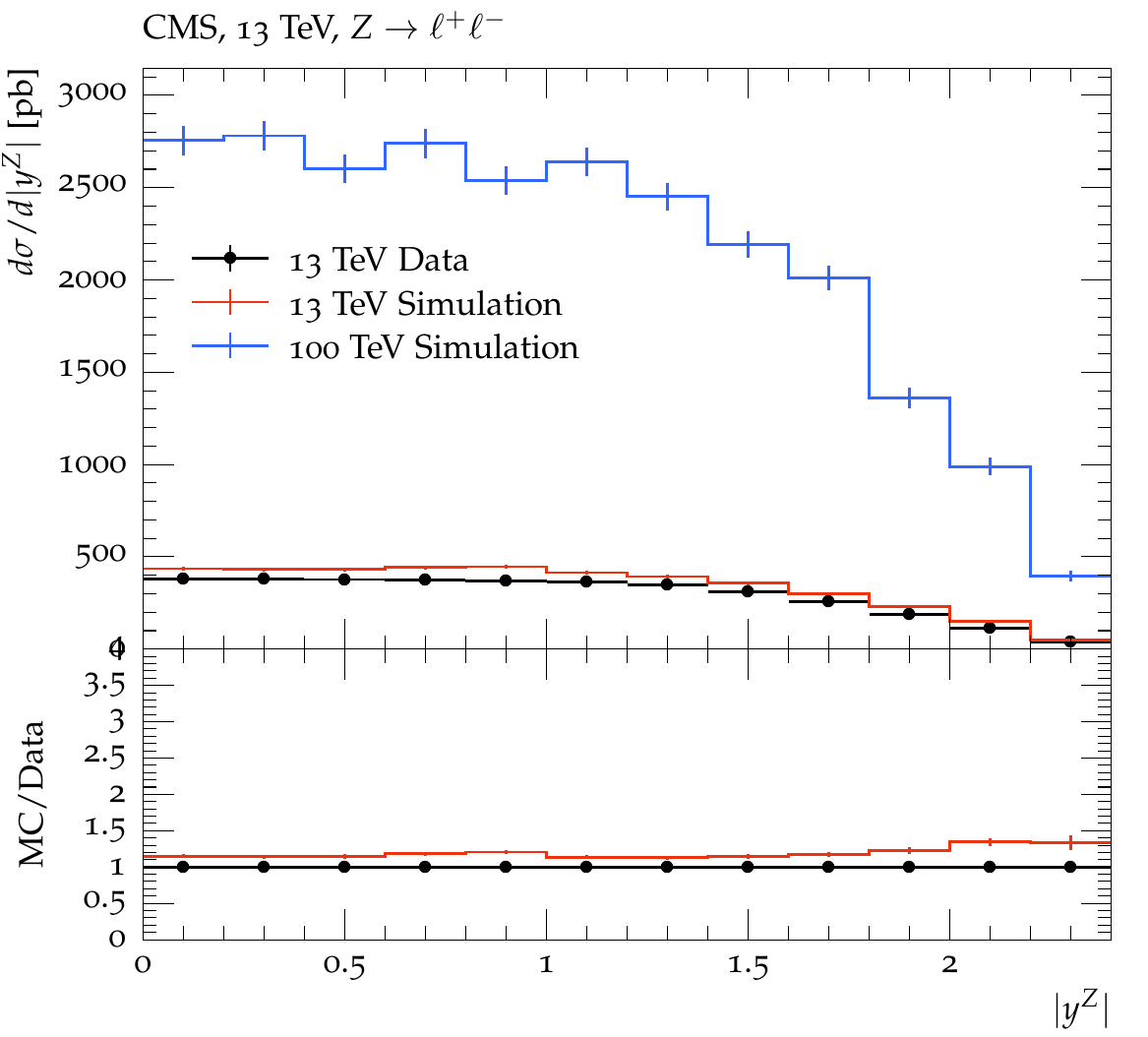}
        \includegraphics[width=0.48\textwidth]{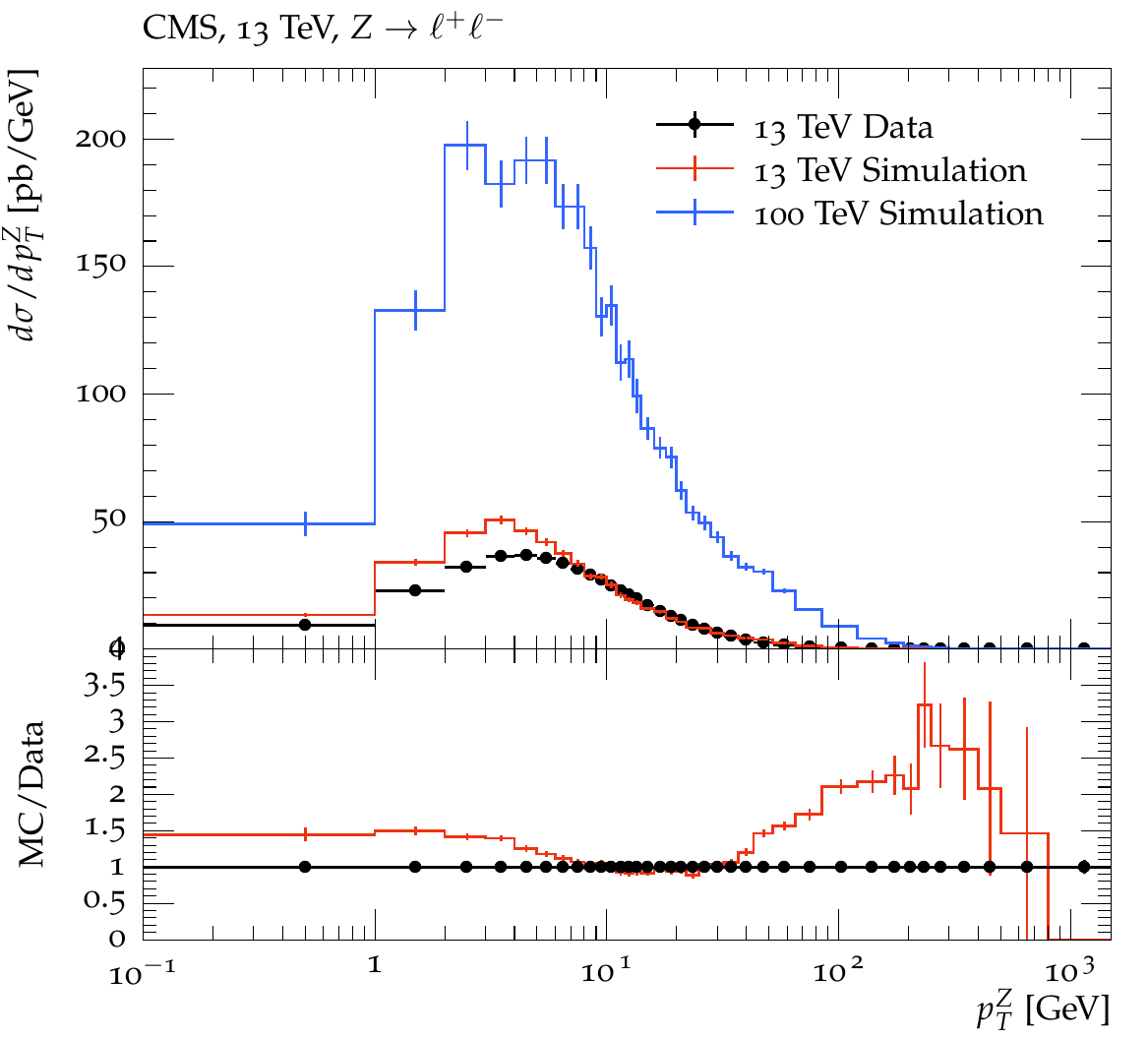}
        \caption{A comparison of $\sqrt{s}=13$\TeV \zplusjets data and simulation, as well as results for $\sqrt{s}=100$\TeV simulation.  The left (right) plot shows the $Z$ boson $y$ ($p_T$).}
    \label{fig:Z}
    \end{center}
\end{figure}

\begin{figure}
    \begin{center}
        \includegraphics[width=0.48\textwidth]{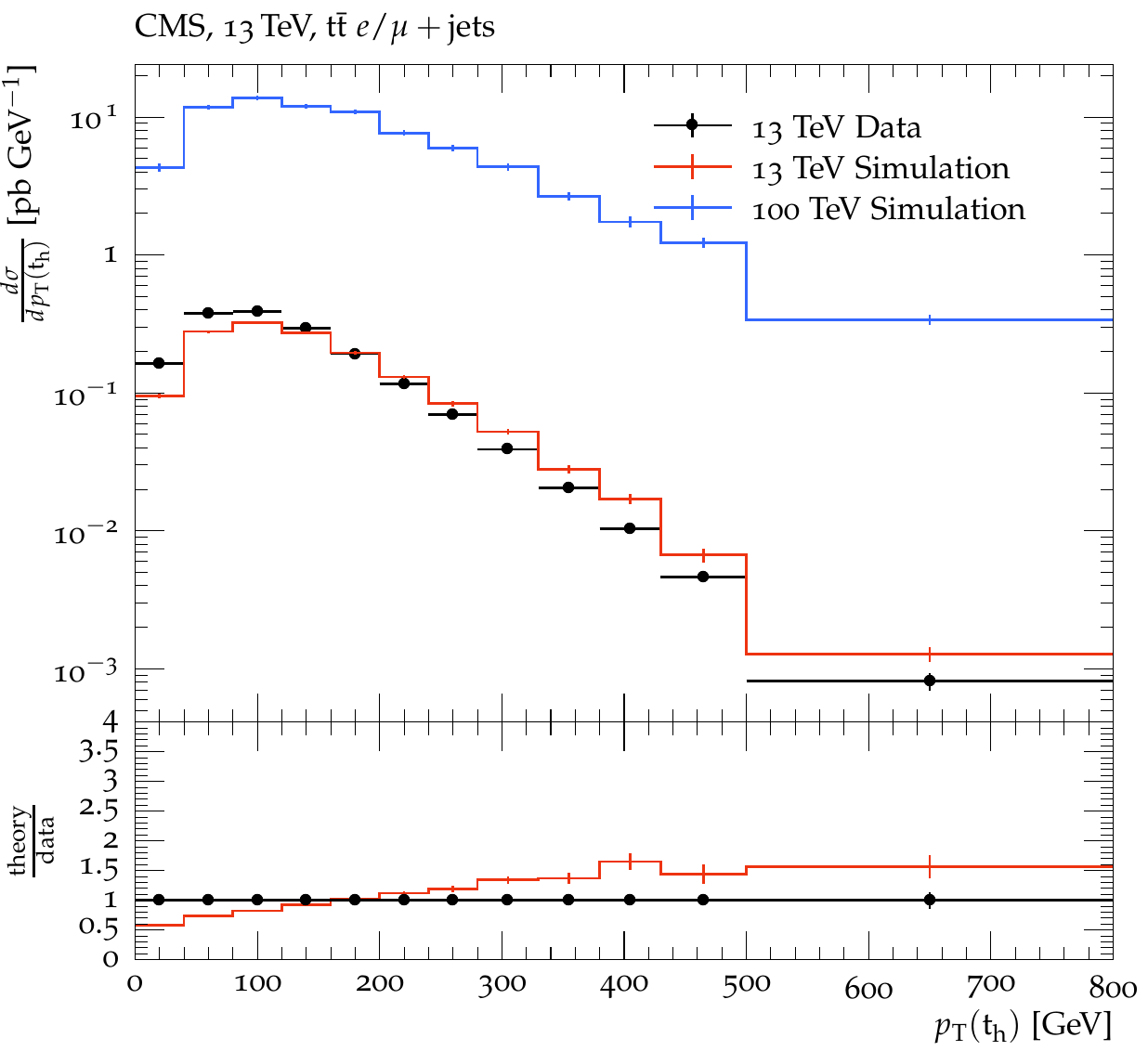}
        \includegraphics[width=0.48\textwidth]{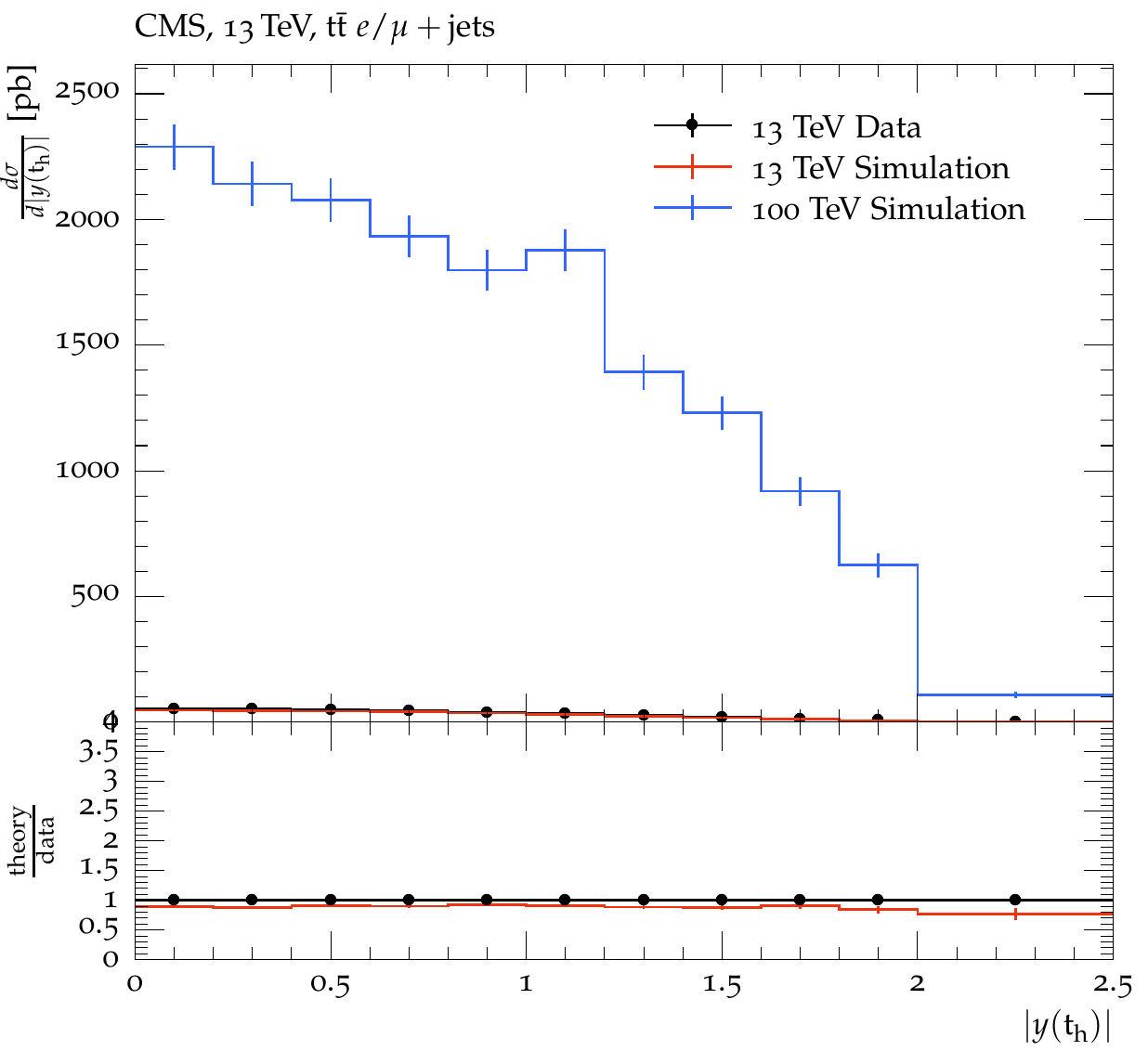}
        \includegraphics[width=0.48\textwidth]{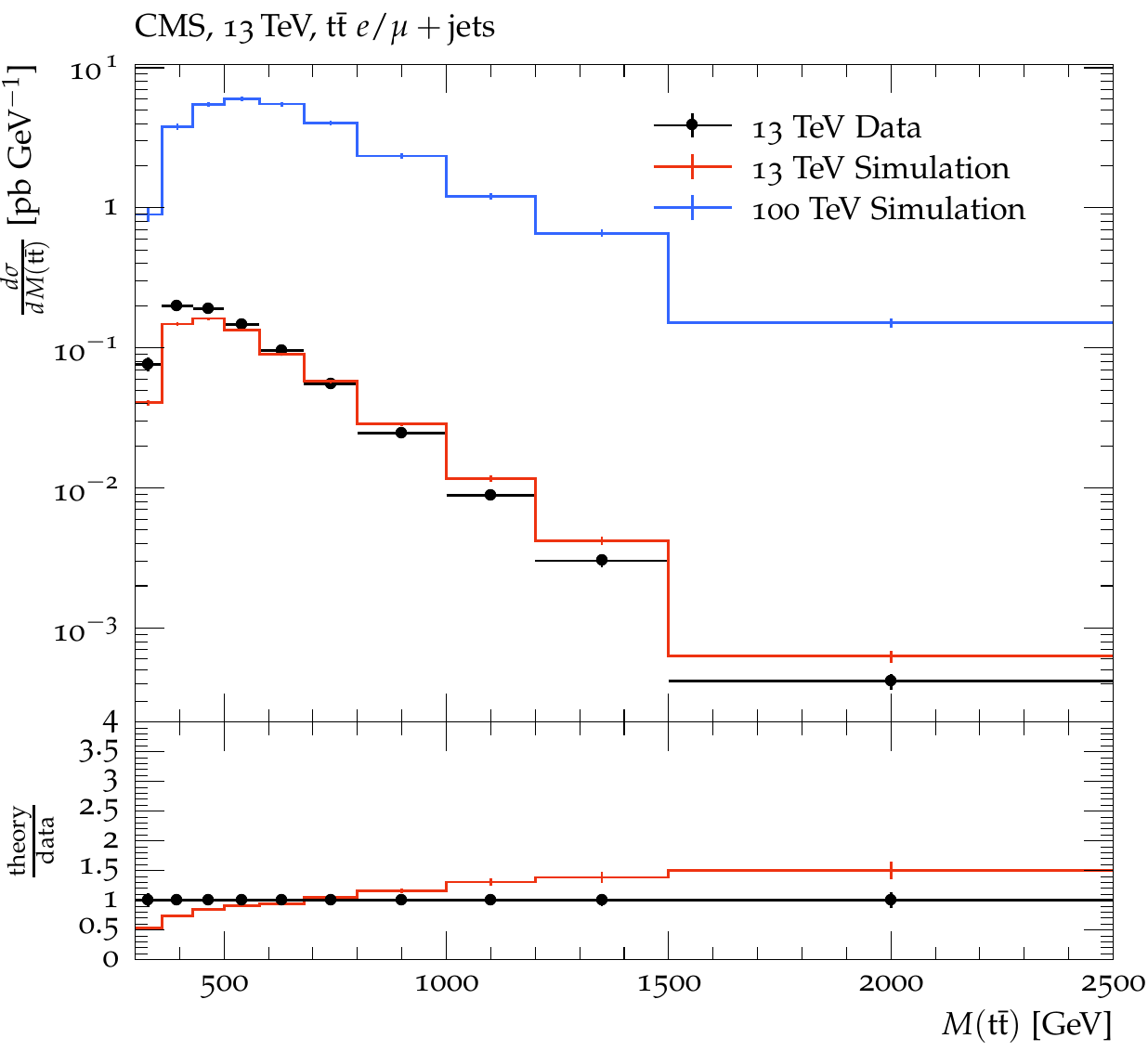}
        \includegraphics[width=0.48\textwidth]{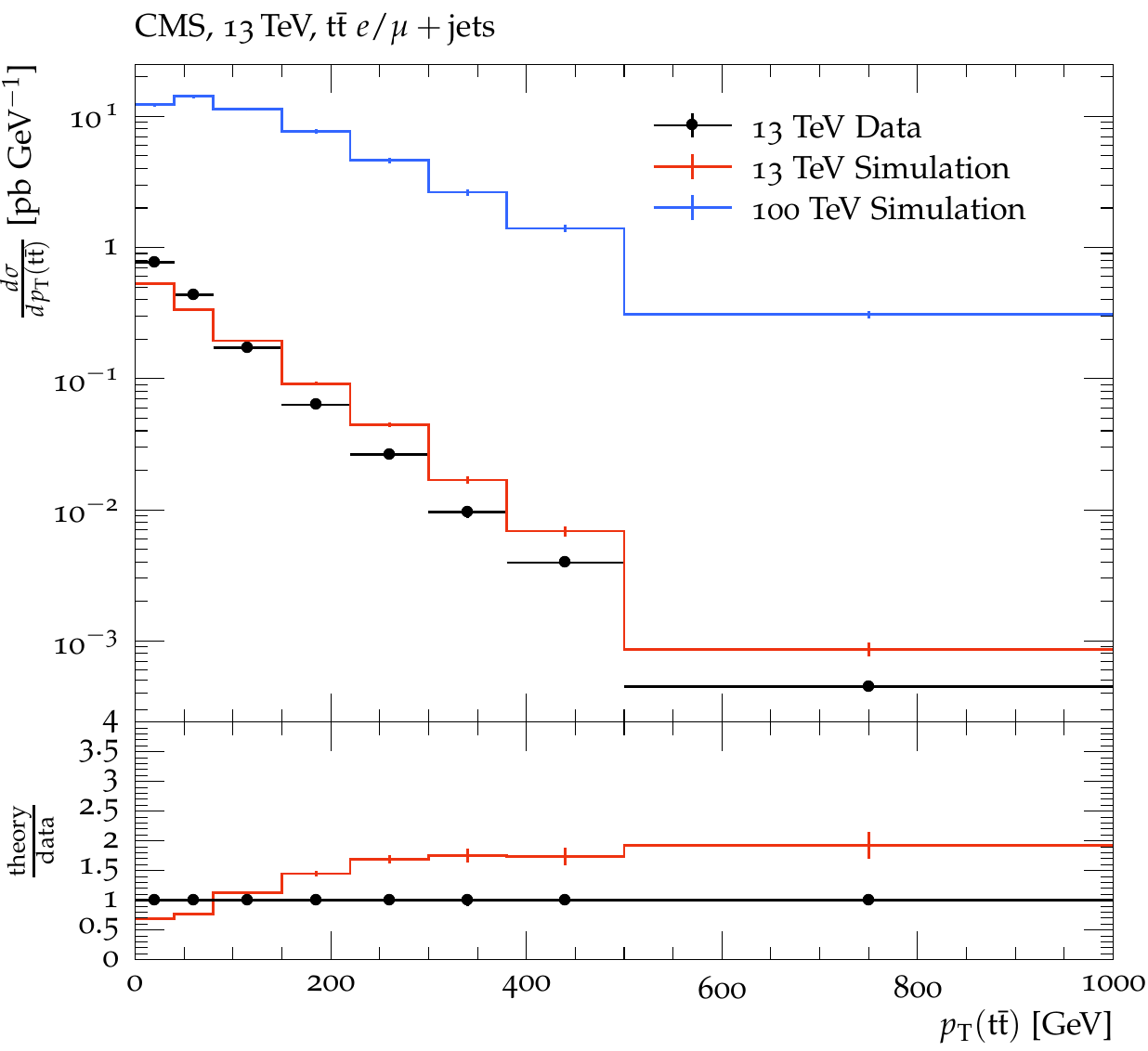}
        \caption{A comparison of $\sqrt{s}=13$\TeV \ttbar data and simulation, as well as results for $\sqrt{s}=100$\TeV simulation.  The upper left (right) plot shows the hadronic top $p_T$ ($y$).  The lower left (right) plot shows the invariant mass $m$ ($p_T$) of the top-pair system.}
        \label{fig:tt}
    \end{center}
\end{figure}

A comparison of the $\sqrt{s}=13$\TeV simulation and data is shown in Figures~\ref{fig:W}, \ref{fig:Z}, and \ref{fig:tt}, for \wplusjets, \zplusjets, and \ttbar samples, respectively. Additional comparison plots are available on the web.\footnote{\wplusjets: \href{https://jstupak.web.cern.ch/noBias/W/}{https://jstupak.web.cern.ch/noBias/W/}, \newline\zplusjets:
\href{https://jstupak.web.cern.ch/noBias/Z/}{https://jstupak.web.cern.ch/noBias/Z/}, \newline\ttbar: \href{https://jstupak.web.cern.ch/noBias/tt/}{https://jstupak.web.cern.ch/noBias/tt/}}  The agreement between simulated \zplusjets and \ttbar samples and data is reasonable;  while the overall discrepancy is approximately 10\%, the shapes are not generally well reproduced in the simulation, leading to bin-by-bin discrepancies of up to 50\% or more in certain regions of phase space.  The agreement between the simulated \wplusjets and data is worse, at the 50\%-level overall.  There is a general overproduction of jets, particularly at high $p_T$.  The source of this discrepancy is under investigation.  However, some event shapes are well described by the simulation.

\section{Conclusion}

The production of SM background MC samples for $\sqrt{s}=13$ and 100\TeV $pp$ colliders has been described.  A comparison of the $\sqrt{s}=13$\TeV simulation to data from the LHC was performed.  While the agreement between simulated \zplusjets and \ttbar events and collision data is reasonable, the modeling of \wplusjets events is not acceptable.  The source of this mismodeling is under investigation.  Due to the similarities in production between the \wplusjets and \zplusjets/\ttbar samples, we do not advise using any of these samples for physics studies, until the discrepancy is understood.


\Acknowledgements{This research was done using services provided by the OSG Consortium \cite{osg07,osg09}, which is supported by the National Science Foundation awards \#2030508 and \#1836650.  We would also like to thank Olivier Mattelaer for significant \madgraph assistance.}

\bibliographystyle{JHEP}
\bibliography{myreferences}  

\appendix

\section{Supplementary Results}

\begin{figure}
    \begin{center}
        \includegraphics[width=0.48\textwidth,page=1,trim=0.9in 4in 0.6in 1.25in,clip]{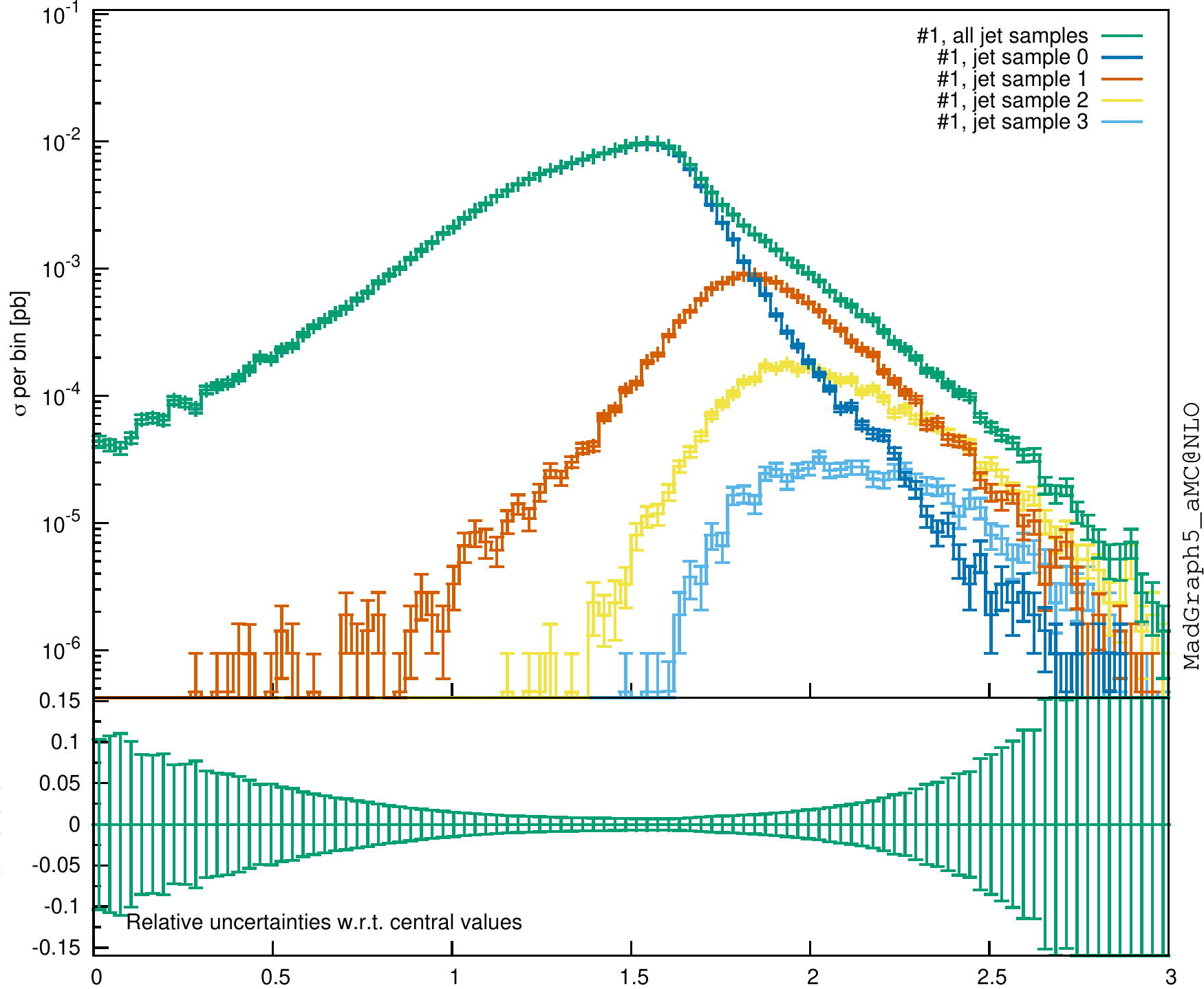}
        \includegraphics[width=0.48\textwidth,page=2,trim=0.9in 4in 0.6in 1.25in,clip]{img/100TeV_Z.tar.pdf}
        \includegraphics[width=0.48\textwidth,page=3,trim=0.9in 4in 0.6in 1.25in,clip]{img/100TeV_Z.tar.pdf}
        \includegraphics[width=0.48\textwidth,page=4,trim=0.9in 4in 0.6in 1.25in,clip]{img/100TeV_Z.tar.pdf}
        \caption{Differential jet rate distributions for $\sqrt{s}=100$\TeV \zplusjets events, showing log$_{10}$ of the merging scale.  The upper left  (right) plot represents a transition from a 0-jet event to a 1-jet (1-jet event to a 2-jet) event.  The lower left (right) plot represents a transition from a 2-jet event to a 3-jet  (3-jet event to a 4-jet) event.}
	\end{center}
\end{figure}

\begin{figure}
    \begin{center}
        \includegraphics[width=0.48\textwidth,page=1,trim=0.9in 4in 0.6in 1.25in,clip]{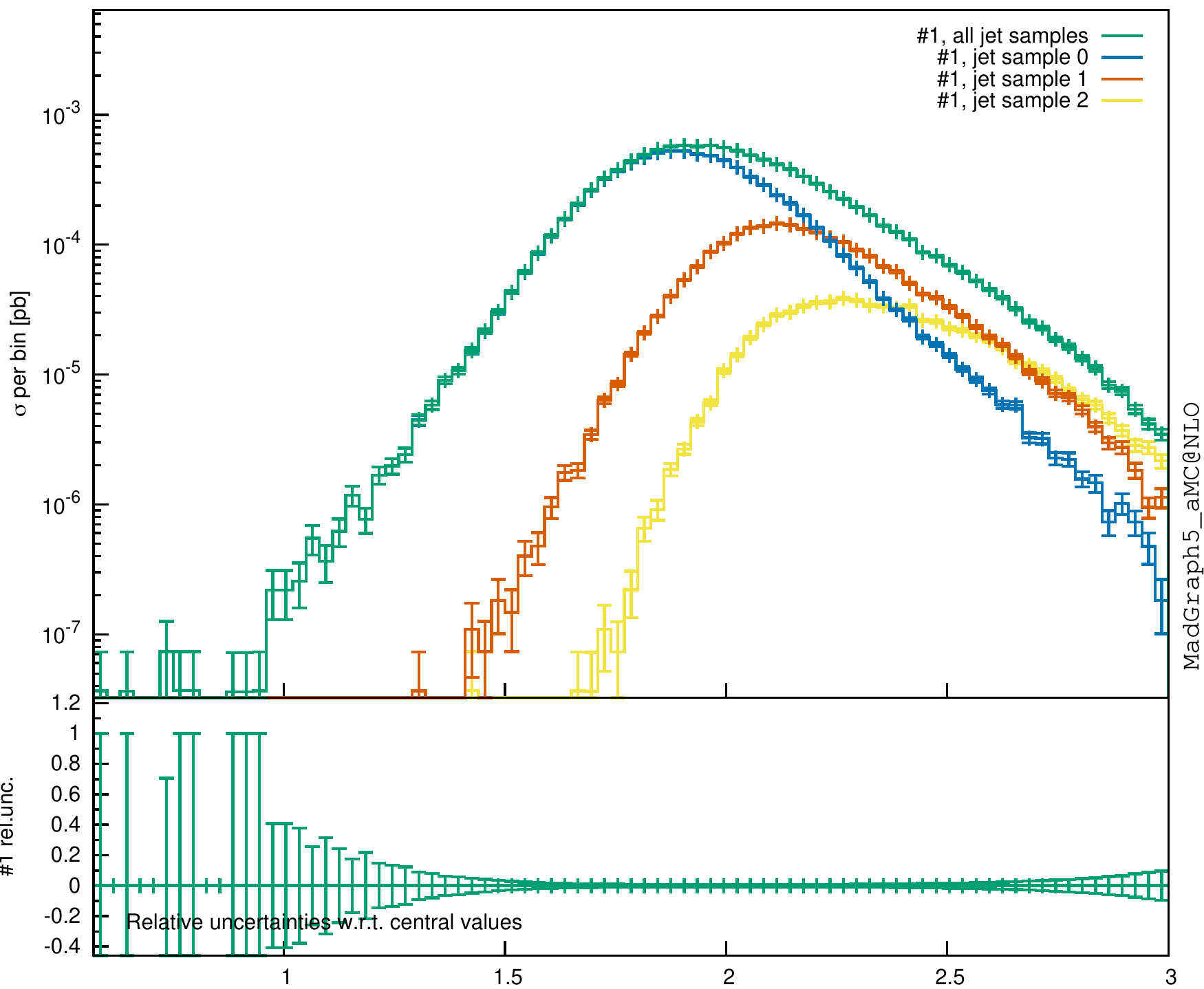}
        \includegraphics[width=0.48\textwidth,page=2,trim=0.9in 4in 0.6in 1.25in,clip]{img/100TeV_tt.tar.pdf}
        \includegraphics[width=0.48\textwidth,page=3,trim=0.9in 4in 0.6in 1.25in,clip]{img/100TeV_tt.tar.pdf}
        \includegraphics[width=0.48\textwidth,page=4,trim=0.9in 4in 0.6in 1.25in,clip]{img/100TeV_tt.tar.pdf}
        \caption{Differential jet rate distributions for $\sqrt{s}=100$\TeV \ttbar events, showing log$_{10}$ of the merging scale.  The upper left  (right) plot represents a transition from a 0-jet event to a 1-jet (1-jet event to a 2-jet) event.  The lower left (right) plot represents a transition from a 2-jet event to a 3-jet  (3-jet event to a 4-jet) event.}
	\end{center}
\end{figure}	

\end{document}